%% file: final.tex
\documentstyle[12pt,epsfig]{article}
\newcommand{\bee}{\begin{equation}}
\newcommand{\ee}{\end{equation}}
\newcommand{\beqn}{\begin{eqnarray}}
\newcommand{\eeqn}{\end{eqnarray}}

\newcounter{savefig}

\newcommand{\NLC}{500 GeV $e^+e^-$ Linear Collider}
\begin{document}
\begin{center}
\boldmath
\Large{\bf
The $\gamma^*\gamma^*$ Total Cross Section and the BFKL Pomeron 
at the 500 GeV $e^{+}e^{-}$ Linear Collider}
\unboldmath
\end{center}
\vspace*{0.5cm}
\begin{center}
Jochen Bartels$^a$, Albert De Roeck$^b$, Carlo Ewerz$^a$, 
Hans Lotter$^{a,}$\footnote{Now at Trinkaus \& Burkhardt KGaA, 
K\"onigsallee 22, D-40212 D\"usseldorf}
\\[.5cm]
$^a$ II.\ Institut f\"ur Theoretische Physik, 
Universit\"at Hamburg,\\ 
Luruper Chaussee 149, D-22761 Hamburg\footnote
        {Supported by Bundesministerium f\"ur Forschung und
        Technologie, Bonn, Germany under Contract 05\,6HH93P(5).}
\\
$^b$ Deutsches Elektronen-Synchrotron DESY, Notkestr.\ 85, D-22603 Hamburg
\end{center}
\vspace*{0.5cm}
 
{\bf Abstract}: We present a numerical estimate of the $\gamma^* \gamma^*$ 
total cross section at the designed 500 GeV $e^+e^-$ Linear Collider,
based upon the BFKL Pomeron. We find that
%, for the linear collider, 
the event rate is substantial
provided electrons scattered under small angles can be detected, 
and a measurement of this cross section provides an 
excellent test of the BFKL Pomeron. 

\section{Introduction}
%{\bf 1.} 
%Recently much attention has been given to the BFKL Pomeron 
%\cite{BFKL}, in particular in the context of small-$x$
%deep inelastic electron proton scattering at HERA. 
In the past years, the BFKL Pomeron \cite{BFKL} 
has been intensively investigated, in particular in the context 
of small--$x$ deep inelastic electron-proton scattering at HERA.  
Whereas the use of this
QCD leading logarithmic approximation for the proton structure function is 
affected by serious theoretical difficulties, it has been argued 
\cite{Hot} that the observation of forward jets near the proton beam 
provides a much more reliable test of the BFKL Pomeron. The main reason for
this lies in the fact that the forward jet 
cross section involves only a single 
large momentum scale, namely the transverse momentum of the forward jet 
which is chosen to be
equal or close to the virtuality of the photon. In the structure function 
$F_2$, on the other hand, the BFKL Pomeron 
feels both the large momentum scale
of the photon mass and the lower factorization scale. In addition, the
diffusion in $\log k_T^2$ always leads to a nonzero contribution of small
transverse momenta where the use of the leading logarithmic approximation
becomes doubtful. A numerical estimate shows that for the forward jets
at HERA \cite{BL} the contribution from this dangerous infrared region is 
reasonably small, whereas in the case of $F_2$ \cite{BLV} the situation is 
much less favorable. As to the experimental situation,  recent analyses of 
HERA data \cite{BDDGW,Woelfle,ADR} on the production of 
forward jets shows a very 
encouraging agreement between data and the theoretical prediction.

In this contribution we would like to point 
out that also $e^+e^-$ linear colliders,
in particular the linear colliders with a rather high luminosity, offer an
excellent opportunity to test the BFKL prediction. 
The process to be looked at is the total cross section of 
$\gamma^* \gamma^*$ scattering. The measurement 
of this cross section requires the double tagging of both outgoing 
leptons close to the forward direction. 
By varying the energy of the tagged leptons 
it is possible to probe the total cross section of the subprocess
$\gamma^* \gamma^*$ from low energies up to almost the full energy of the 
$e^+ e^-$ collider. 
For sufficiently large photon virtualities we again have 
a situation with only large momentum scales. In other words, photons with
large virtualities are objects with small transverse size, and it is
exactly this situation for which the BFKL approximation should be 
considered most reliable. 
The energy dependence of this cross section, therefore,
should be described by the power law of the BFKL Pomeron. 
%%%%%%%%%%%%%%%%
\begin{figure}[htbp]
\begin{center}
\input gamma.pstex_t
\caption{Feynman diagram for the process 
$e^+e^- \to e^+e^- + \mbox{anything}$.}
\end{center}
\end{figure}
%%%%%%%%%%%%%%%%

\noindent
{From} the theoretical point of view it 
is clear that we want the photon masses 
to be large. On the other hand, because of the photon propagators, the 
$e^+e^-$ cross section for this final state (i.\ e.\ the event rate for the
process under discussion) falls off very rapidly with increasing photon 
masses. Therefore, one cannot afford to have too large photon virtualities. 
As a compromise, we chose the range of $5$ to 
$200 \,\mbox{GeV}^2$ (for 
experimental considerations see further below). As to the energies of the 
$\gamma^*\gamma^*$ subprocess, we can 
in principle go up to almost the full 
collider energies. From the theoretical side, however, 
it is important to estimate the diffusion in the internal 
transverse momenta
$k_T$. In the center of the BFKL ladders (Fig.1), the
distribution in $\log k_T^2$ is given by a Gaussian, with center at
$\log Q^2$ (if we chose, for simplicity, 
both photon virtualities to be equal),
and with the width growing linearly with 
the square root of rapidity. 
As soon as the
small-$k_T$ part of the Gaussian reaches the confinement region
the BFKL prediction (which is based upon a leading--log calculation) 
becomes unreliable. 
Corrections to the BFKL Pomeron, in particular those which are expected
to restore unitarity are no longer small.
Qualitatively they are expected to reduce the growth
of the cross section with increasing energy.
%%%%%%%%%%%%%%%%%
Below we will argue that 
this energy region can be reached at the linear collider. 
In other words, at highest energies for the $\gamma^*\gamma^*$ 
subprocess, a deviation from the power--rise of the BFKL Pomeron might 
become visible.

This note is based mainly on \cite{BDL}. 
Here we give a more detailed analysis of event rates 
and their dependence on experimental resolution.
Moreover, we propose a more refined way to extract 
a clean BFKL signal from the data. This is done by 
considering event samples in which the virtualities of the 
two photons are of the same order. We introduce a parameter 
that measures how much the virtualities differ. This 
parameter is easily accessible experimentally, and we give 
expected event rates for different parameter values. 

The investigation of the $\gamma^*\gamma^*$ total cross 
section as a probe of BFKL dynamics 
has independently been advocated by Brodsky et al.\ \cite{Brodsky}. 
The cross section formulae obtained there are in perfect 
agreement with our results presented here in section 2.
The cross section has also been discussed by Bialas et al.\ 
within the framework of the colour dipole picture of the 
BFKL Pomeron \cite{Bialas}.

\section{Cross Section Formulae}
%{\bf 2.} 
The theoretical prediction of the cross section is based on  the high 
energy behaviour of the diagrams shown in Fig.\ 1. 
Let us first define suitable variables. 
In analogy to DIS kinematics we chose the scaling variables
\beqn
y_1=\frac{q_1 k_2}{k_1k_2}, \,\,\, y_2=\frac{q_2k_1}{k_1k_2} 
\eeqn 
and
\beqn
x_1=\frac{Q_1^2}{2q_1k_2},\,\,\, x_2=\frac{Q_2^2}{2q_2k_1}
\eeqn
where the photon virtualities are, as usual, $Q_i^2=-q_i^2$ ($i=1,2$). 
Energies are denoted by $s=(k_1+k_2)^2$ and 
$\hat{s}=(q_1+q_2)^2 \approx sy_1y_2$. With our definitions of the scaling 
variables we have $Q_i^2=sx_iy_i$ ($i=1,2$). We consider the limit of large
$Q_1^2$, $Q_2^2$, and $\hat{s}$ with 
\beqn
Q_1^2,\,Q_2^2 \ll \hat{s}.
\eeqn 
The calculation is straightforward and,
neglecting terms of the order of $Q_i^2/\hat{s}$,
leads to the following result:
\beqn
\frac{d \sigma^{e^+e^-} }
{d Q_1^2 d Q_2^2 dy_1 dy_2}
&=& \frac{\alpha^2_{em}}{16 \pi y_1 Q_1^4}
  \frac{\alpha^2_{em}}{16 \pi y_2 Q_2^4}
\int \frac{d \nu}{2 \pi^2} \; \exp\left[\log s 
\frac{y_1 y_2}{\sqrt{Q_1^2 Q_2^2}} \cdot \chi(\nu) \right] 
\times
\nonumber \\
& &
%\!\!\!\!\!\!\!\!\!\!\!\!\!\!\!\!\!\!\!\!\!\!\!\!\!\!
%\!\!\!\!\!\!\!\!\!\!\!\!\!\!\!\!\!\!\!\!\!\!\!\!\!\!\!
\times
\left[(1-y_1)W^{(1)}_L(\nu)+\frac{1+(1-y_1)^2}{2}W^{(1)}_T(\nu)\right]
\nonumber\\
& &
\times
\left[(1-y_2)W^{(2)}_L(-\nu)+\frac{1+(1-y_2)^2}{2}W^{(2)}_T(-\nu)\right] 
\label{4}
\eeqn
with 
$\chi(\nu)= (N_c \alpha_s/\pi) [2\psi(1)-\psi(1/2+i\nu)-\psi(1/2-i\nu)]$
and we have introduced the invariant functions:
\beqn
W_L^{(i)}(\nu)&=& 
\sum_f q_f^2 \alpha_s \;\pi^2 
\sqrt{2} \; 8  
\frac{\nu^2+\frac{1}{4}}{\nu^2+1} 
\frac{\sinh \pi \nu}{\nu \cosh^2 \pi \nu}
(Q_i^2)^{\frac{1}{2}+i\nu}
\label{5}
\\
W_T^{(i)}(\nu)&=&
\sum_f q_f^2 \alpha_s \;\pi^2 
\sqrt{2} \; 4
\frac{\nu^2+\frac{9}{4}}{\nu^2+1} 
\frac{\sinh \pi \nu}{\nu \cosh^2 \pi \nu}
(Q_i^2)^{\frac{1}{2}+i\nu}
\label{6}
\eeqn
Here $q_f$ is the quark charge, and in our calculations the sum 
over (massless) flavours goes up to four. For the scale of the 
%first order 
strong coupling constant we use $\sqrt{Q_1^2 Q_2^2}$
($\alpha_s(M_Z^2)=0.12$).

Before we turn to the discussion of numerical results, let us estimate the
diffusion of internal transverse momenta into the infrared region. In the 
center of the BFKL ladders (Fig.\ 1), the width of 
the Gaussian distribution of
$\log k_T^2$ is given by \cite{BL}
\beqn
\Delta = \sqrt{<(\log k_T^2 - <\log k_T^2>)^2>}  
       =   \sqrt{7 \frac{N_c \alpha_s}{\pi} \zeta(3) \log s/s_0} 
%       = \mbox{const.} \cdot \sqrt{ \log s/s_0}
\eeqn
with 
\beqn
 s_0 = \frac{\sqrt{Q_1^2 Q_2^2}}{y_1y_2} \,. 
\eeqn
For a typical value $Q_1^2 = Q_2^2=10 \,\mbox{GeV}^2$, and 
$\alpha_s=0.22$ one finds, for the maximal value $Y=\log s/s_0 =10$ at the 
\NLC, $\Delta=4.2$ which means that the diffusion reaches down to $k_T^2 
\approx 0.15$ GeV$^2$. 
%For LEP the maximal value for $Y$ is near 6, and
%we obtain $\Delta=3.2$ and $k_T^2=0.4 \,\mbox{GeV}^2$. 
Compared to the forward jets at HERA 
\cite{BL} where the corresponding value lies 
above $1 \,\mbox{GeV}^2$, we now have to 
expect a substantially larger contribution from the small-$k_T$ region: this
should lead to a lowering of the BFKL power behaviour of the cross section.
In other words, one might be able to see the onset of unitarity corrections 
to the BFKL Pomeron. 

\boldmath
\section{Cross Section at the 500 GeV $e^+e^-$ Linear Collider}
\unboldmath
%{\bf 3.} 
Starting from eqs.\ (\ref{4})--(\ref{6}) 
we have calculated the differential 
cross section for different values of the logarithm of the subenergy 
$Y=\log s/s_0$.
In order to illustrate
the BFKL power law, we have multiplied the cross section by $y_1y_2$.
In Fig.\ 2 we show the results for 
the $500$ GeV Next Linear Collider.
On the left hand side we display the $e^+e^-$ cross section
for $Q_1^2=Q_2^2=10 \,\mbox{GeV}^2$. 
The right hand side shows the cross section for 
$Q_1^2=Q_2^2=25 \,\mbox{GeV}^2$. 
For comparison, we have plotted on the left in each figure 
the corresponding cross sections at LEP I (91 GeV $e^+e^-$). Due to the 
larger energy, the cross sections at the \NLC\ are considerably 
larger. The much higher design luminosity of 
the Linear Collider will lead to 
an even larger difference in event rates.
%%%%%%%%%%%%%%%%%%%%%%%%%%%
\begin{figure}[ht]
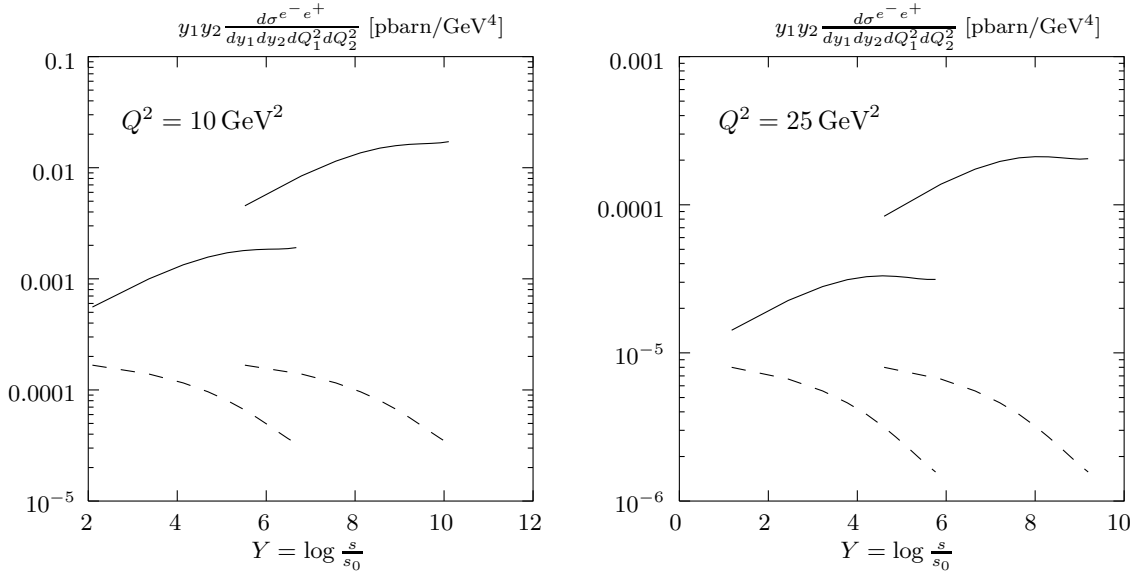

\begin{center}
\input fig2.pstex_t
\caption{
The differential 
$e^+e^-$ total cross section, multiplied by $y_1y_2$, as a function
of the rapidity $Y=\log s/s_0$. The full curves denote the 
%exact 
numerical 
calculation based upon eq.\ (\ref{4}) 
and the dashed lines represent the 
2--gluon approximation (no gluon production between the two quark pairs). 
The two curves on the left in each figure 
 are for  LEP I, the ones on the right for the \NLC. 
We have chosen $y_1=y_2$ and $Q_1^2=Q_2^2 = 10 \,\mbox{GeV}^2$
(left hand figure) resp.\ $Q_1^2=Q_2^2 = 25 \,\mbox{GeV}^2$
(right hand figure).
}
\end{center}
\end{figure}
%%%%%%%%%%%%%%%%%%%%%%%%%%%
The full curves represent the results 
based upon (\ref{4})--(\ref{6}). 
The dashed lines denote the process in which there is no 
gluon production between the two fermion pairs (2--gluon 
approximation). This process is the one of lowest order 
in $\alpha_s$ that gives a sizable contribution to the
cross section, since the quark--box 
diagram is strongly suppressed at high energy and can 
be neglected. Therefore, we compare our BFKL resummed 
prediction with this fixed order cross section. 

From the curves in Fig.\ 2 one recognizes the typical 
BFKL power-like energy behaviour, but there 
is some damping at large rapidity (large $y_i$) due to the photon flux 
factors in (\ref{4}).
In the 2--gluon cross section this effect even leads to a decrease
at large rapidity $Y$. The BFKL predictions are 
well above the 2--gluon curves, up to more than an order of magnitude.
Comparing the right hand side of Fig.\ 2 with the left hand side 
we find that by increasing $Q^2$ from 10 to 25 $\mbox{GeV}^2$
the cross section decreases by two orders of magnitude.
As seen from (\ref{4}), 
the cross section scales with $1/Q^6$, and there is
an additional decrease at larger $Q^2$ due to the $Q^2$-dependence of 
$\alpha_s(Q^2)$. 
%%%%%%%%%%%%%%%%%%%%%%%%%%%
\begin{figure}[htbp]
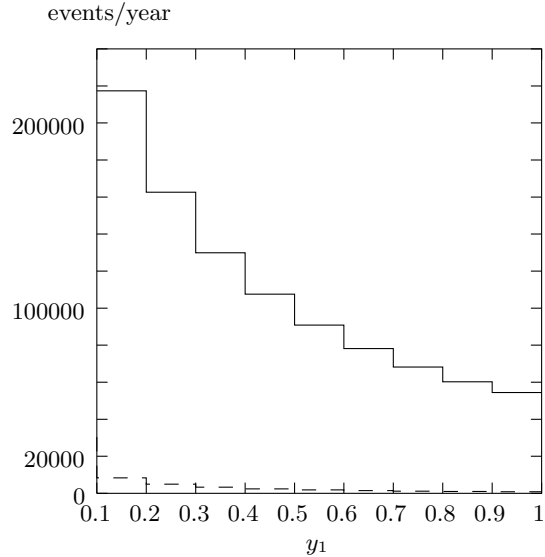

\begin{center}
\input fig3.pstex_t
\caption{
The total number of events per year ($3\cdot 10^7 \mbox{s}$), 
as a function of $y_1$. 
The variables
$Q_i^2$ ($5<Q_i^2<200$ GeV$^2$) and $y_2$ ($0.1<y_2<0.9$) are integrated.
%The left hand side corresponds to the LEP and the right hand side to 
%the NLC situation.
The solid curve shows the results of the BFKL calculation 
%the dashed curve shows the approximation 
and the dashed curve 
represents the 2--gluon result.
}
\end{center}
\end{figure}
%%%%%%%%%%%%%%%%%%%%%%%%%%%
Event rates for the \NLC\ are shown in Fig.\ 3, taking 
$3\cdot 10^7$ s/year and assuming a 
luminosity of ${\cal L}=10^{33} \,\mbox{cm}^{-2} \mbox{s}^{-1}$. 
Due to the limit in time of data taking periods, 
experiment and accelerator efficiency, this corresponds effectively 
to several years of operation of the accelerator.
%In the following, we will therefore give event rates as 
%expected for an integrated luminosity of $1000 \,\mbox{pb}^{-1}$. 
In $Q_1^2$ and $Q_2^2$ we integrate from $5$ to 
$200 \,\mbox{GeV}^2$, and for the
$y_1$-variable we have chosen 9 bins, as indicated in the figure. The other
$y$-variable is integrated from $0.1$ to $1.0$. The rapidity
of the subprocess $\gamma^* \gamma^*$ is restricted by $\log s/s_0 > 2$.
The event rate calculations are based upon Monte Carlo integration 
of the phase space, and the accuracy is of the order of 5\%.  

\section{Experimental Restrictions and Event Rates}
%{\bf 4.} 
Due to experimental restrictions, however, these event rates
can only give a first impression and not more.
The measurement of the total cross section of 
$\gamma^* \gamma^*$ scattering 
can be made at existing and future $e^+e^-$ colliders
using so called ``double tag'' events. These are events where both 
outgoing leptons are detected and some hadronic 
activity is observed in the central detector.
The $Q^2$ value of the virtual photon emitted from the lepton is
$Q^2= 2E_bE_{tag}(1-\cos\theta_{tag})=
4E_bE_{tag} \sin^2\frac{\theta_{tag}}{2}$, 
with $E_{tag}$ and 
$\theta_{tag}$ the energy and angle of the tagged lepton, 
and $E_b$ the energy of the incident lepton.
The variable $y$ is given 
by $y=1-(E_{tag}/E_b)\cos^2\frac{\theta_{tag}}{2}$.
Combination of the two relations leads to the convenient equation
\beqn
Q^2=4E_b^2(1-y)\tan^2\frac{\theta_{tag}}{2}
\label{11}
\eeqn
which holds for any of the two incoming leptons.
Experiments at LEP tag electrons down to about 25 mrad \cite{opal}
leading to $Q^2$ values as low as 
to $1 \,\mbox{GeV}^2$. 
For  $Q^2$ values of 5 GeV$^2$ the electrons are scattered under 
an angle of 60 mrad, well
within the forward detector acceptance of the LEP experiments.
In order to reach such 
$Q^2$ values at a \NLC, the scattered leptons need to be detected down 
to 10 mrad. For photons of virtuality $20 \,\mbox{GeV}^2$  
angles down to about 15--20 mrad need 
to be covered. 
According to the conceptual design report~\cite{NLC}, the acceptance of the 
forward endcap electromagnetic calorimeters is foreseen to go down
to angles of 80 mrad,
 leading to minimum reachable 
$Q^2$ values in the range of $60-80 \,\mbox{GeV}^2$. 

One of the main problems in going down to smaller angles 
at the \NLC\ will be  $e^+e^-$
pair production.
At the interaction point beam-beam effects cause production 
of beamstrahlung, which is very intense and collimated in a small cone
around the beam axis. It will increase the background of $e^+e^-$
pairs produced by two colliding real photons, which has been studied 
in~\cite{schulte}. The studies indicate that tagging of scattered
electrons 
down to angles of 30 mrad is within reach. 
These  electrons could be detected with the proposed small angle 
calorimeters,
designed to measure the luminosity, with acceptance in the 
region $30-55$ mrad.

The $y$ values reached in present single tagged analysis at LEP 
are in the range
$y<0.25$. However using double tagged events the background should 
be kept well under control also for larger $y$ values, and therefore 
values of $y=0.5$ or more, which lead to a large mass system for the 
hadronic final state and, correspondingly, to an extended ladder, are a 
realistic goal. Note that for all calculations presented here 
 no requirement on the detection of the hadronic final state
was made. Experimentally
it is likely that some (very weak)  cuts will have to be applied to 
discriminate the signal from background, which will reduce the rates 
somewhat.

%%%%%%%%%%%%%%%%%%%%%%%%%%%
\begin{figure}[htbp]
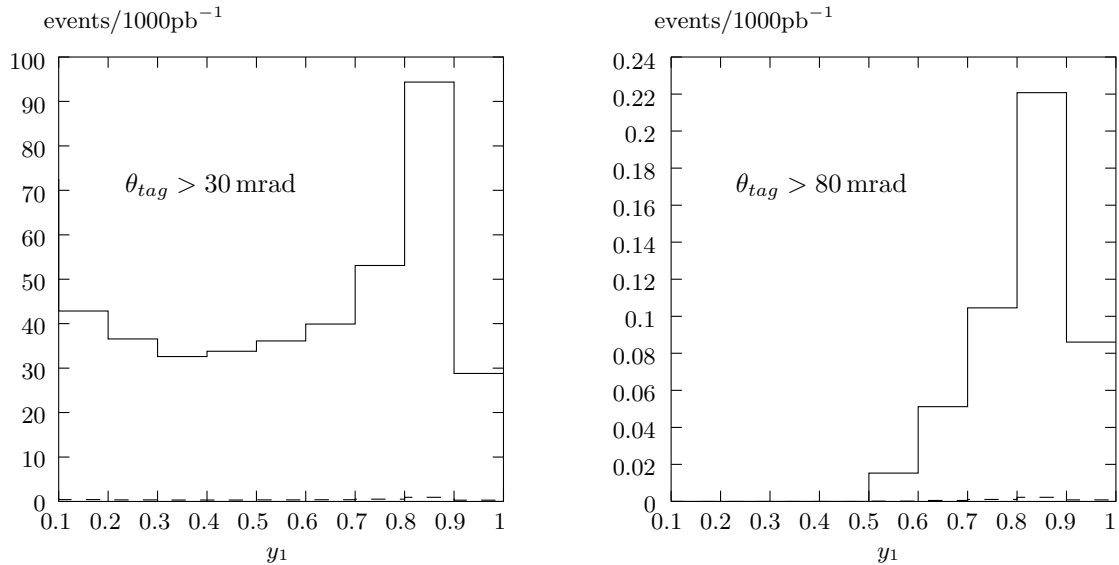

\begin{center}
\input fig4.pstex_t
\caption{
The total 
number of events per 1000 $\mbox{pb}^{-1}$ 
with detector cuts in angle taken into account.
This would typically correspond to one month of data taking at the 
design luminosity. 
We have chosen $E_{tag}>20$ GeV, $\log s/s_0>2$, 
$2.5 <Q_i^2 <200$ GeV$^2$.
The acceptance cuts are $\theta_{tag}> 30 \,\mbox{mrad}$ (left) 
and $\theta_{tag}> 80 \,\mbox{mrad}$ (right). 
The $y$--binning is the same as in fig.\ 3. Solid lines are again 
the prediction based on BFKL, dashed lines represent 
the 2--gluon exchange. 
}
\end{center}
\end{figure}
In order to incorporate these conditions into our calculation of 
event rates for the \NLC, 
we have repeated the previous calculations with more 
realistic kinematical cuts. 
We impose the constraints 
$\theta_{tag}>30$ mrad (Fig.\ 4, left) and $\theta_{tag}>80$ mrad
(Fig.\ 4, right), and
we require $E_{tag}>20$ GeV (i.\ e.\ approximately 
$y_i<0.9$). 
Further restrictions are $2.5<Q_i^2<200 \,\mbox{GeV}^2$ and
$\log s/s_0>2$.
In particular for smaller $y_1$ the event rate is substantially 
lower than the corresponding one in Fig.\ 3. (Fig.\ 3 shows 
event rates corresponding to 30000 $\mbox{pb}^{-1}$.
To compare with numbers there, we thus have to multiply numbers in 
Fig.\ 4 by 30.)
This shows the importance of the angle cut. 
Clearly, $\theta_{tag}=30 \,\mbox{mrad}$ 
looks highly desirable. 
For $\theta_{tag}=80$ mrad, 
the $Q^2$ values at the \NLC\ are rather large, reducing very strongly
the event rate.
The suppression of the small-$y$ region in the right hand diagram 
of Fig.\ 4 results from the upper limit of the $Q^2$ range taken.

Further, we have computed a few integrated event rates. 
The results are shown 
in Table 1. 
\begin{table}[htbp]
\begin{center}
\renewcommand{\arraystretch}{1.2}
\begin{tabular}{|r||c|c|} \hline
 $\theta_{tag,min}$ & BFKL & 2--gluon \\ \hline 
 20 mrad  &   2724 &   46    \\ \hline
 30 mrad  &    324 &   12    \\ \hline
 40 mrad  &     67 &    3.6   \\ \hline
 50 mrad  &     19 &    1.1   \\ \hline
 60 mrad  &    4.8 &    0.24   \\ \hline
 70 mrad  &    1.3 &    0.059   \\ \hline
 80 mrad  &   0.46 &    0.020   \\ \hline
 90 mrad  &   0.17 &    0.0074   \\ \hline
 100 mrad &  0.063 &    0.0029 \\ \hline
\end{tabular} 
\caption{
Total event rates per $1000 \,\mbox{pb}^{-1}$ for different 
values of $\theta_{tag,min}$. 
We have chosen $E_{tag}>20$ GeV, $\log s/s_0>2$, 
$2.5 <Q_i^2 <200$ GeV$^2$.
}
\renewcommand{\arraystretch}{1}
\end{center}
\end{table}
%
%%%%%%%%%%%%%%%%%%
We have integrated $2.5<Q_i^2<200\,\mbox{GeV}^2$, 
$0.1<y_i<1.0$, with the
constraints $E_{tag}>20$ GeV, 
$\log s/s_0>2$, and for the detector angles
we have chosen $\theta_{tag}>20 - 100$ mrad.
For the smallest angles 
the ratio of BFKL to 2--gluon cross section is around 60.
Note that according to the present design,
 the  detector will not 
measure 
electrons with angles between 55 and 80 mrad, due to the 
mask to shield against background.
For angles above 80 mrad 
%for which single tag measurements at LEP exist,
the BFKL cross section is by a 
factor of 23 larger than the 2--gluon cross section,
but the limitation in $Q^2$ strongly affects the rate as discussed above. 
%The total number of events produced in the four LEP experiments 
%together at LEP-90, taking a total collected integrated luminosity
%of 150 pb$^{-1}$/experiment is twice the number of events given in Table 1.
%Reversely, the LEP-180 numbers correspond to a total number of events
%collected by the experiments if 75 pb$^{-1}$ will be 
%delivered by the machine.
%The experiments at LEP-180 are foreseen to measure already now 
%down to angles about 30 mrad. 
%Table 1 shows that the event rate for this angle
%is reasonable.
From this we conclude that, 
mainly because of the high luminosity of the \NLC,
such a machine offers an excellent possibility 
to observe the BFKL Pomeron, provided the detectors 
can reach
small angles 
down to about $20-30$ mrad. 
%For LEP the total event rate looks less encouraging,
%but it is still worthwhile to pursue further studies in this direction.

So far, we have calculated event rates applying only the cuts 
$2.5 <Q_i^2 <200$ GeV$^2$ for each of the photons. 
Strictly speaking, the BFKL calculation is applicable only 
for $Q_1^2/Q_2^2 = {\cal O}(1)$. For this kinematic region 
there is no DGLAP evolution between the two quark loops, 
and the BFKL results have to be compared to the 2--gluon 
exchange. For the full region $2.5 <Q_i^2 <200$ GeV$^2$ 
this may be a rather crude approximation, and in a future 
step our analysis will have to be refined by using a 
background calculation based upon DGLAP evolution. 
For the moment, we propose to circumvent this difficulty 
by introducing an additional cut parameter $\rho$ limiting 
the ratio of the two virtualities, 
\beqn
  \frac{1}{\rho} < \frac{Q_1^2}{Q_2^2} < \rho \,.
\label{qcut}
\eeqn
This cut can easily be applied experimentally because we consider 
double tagged events. 
The closer to $1$ the value chosen for $\rho$ is the 
cleaner will be the BFKL sample obtained. 
Lowering $\rho$ of course restricts phase space and thus lowers 
the number of events expected. The high luminosity of the \NLC\ 
will nevertheless make a value for $\rho$ reasonably 
close to $1$ affordable. 

In Table 2 we present event rates for an acceptence 
angle of $\theta_{tag}>30$ mrad, an integrated 
luminosity of $1000 \,\mbox{pb}^{-1}$, and different values 
of $\rho$. We have also varied the tagging energy of the electron 
$E_{tag}$ to larger, i.\ e.\ less optimistic, values. 
The value $\rho = \infty$ corresponds to 
removing the additional cut and leads to the event rates discussed 
above. 
\begin{table}[htbp]
\begin{center}
\renewcommand{\arraystretch}{1.2}
\begin{tabular}{|l|c|c|c|} \cline{2-4}
 \multicolumn{1}{c|}{} & $E_{tag} > 20 \,\mbox{GeV}$ 
   & $E_{tag} > 50 \,\mbox{GeV}$
   & $E_{tag} > 100 \,\mbox{GeV}$\\ 
\hline  
$\rho = \infty$ & 324 (12) & 159 (9) & 72 (6) \\
$\rho = 3$      & 188 (9)  & 122 (7) & 63 (5)\\
$\rho = 2$      & 122 (6)  & 87 (5)  & 47 (4)\\ 
$\rho = 1.5$    & 65 (4)   & 51 (3)  & 31 (3)   \\ 
\hline  
\end{tabular} 
\caption{Total event rates for $1000 \,\mbox{pb}^{-1}$ and different 
values of $E_{tag}$ and $\rho$. We have chosen 
$\theta_{tag}>30 \,\mbox{mrad}$, the other cuts are as above. 
Given are numbers of events for 
BFKL and for 2--gluon exchange (in brackets).
}
\renewcommand{\arraystretch}{1}
\end{center}
\end{table}
Having in mind the high luminosity of the \NLC, the numbers look 
very promising even for the smaller values of $\rho$. 
The cross section based upon the BFKL calculation 
is still a factor of about 15 larger 
compared to the 2-gluon cross section. Requiring 
larger $E_{tag}$ values reduces the cross section considerably, but the 
event rate remains large enough to allow for a study of the BFKL Pomeron
at the \NLC.

\section{Summary}
%{\bf 5.} 
In summary, we have estimated the total cross section of
$\gamma^* \gamma^*$ scattering at 
%both LEP and the  NLC.
the designed 500 GeV $e^+e^-$ Linear Collider. 
At high energies, this subprocess is dominated by the BFKL Pomeron and 
therefore provides an ideal test of this QCD calculation. 
A rough estimate of 
the diffusion in $\log k_T^2$ shows that, at the high energy tail of this 
subprocess, corrections to the BFKL become non-negligible, and hence 
deviations from the BFKL power law may become visible.

For a realistic estimate of the number of observable events we 
find a strong dependence on detector restrictions. 
In order to have a sufficiently large 
number of events, especially when applying the additional 
requirement that the photon virtualities are approximately equal, 
it is necessary to tag both leptons close to the beam 
direction. An angle of 30 mrad might already lead to good results, 
even better would be 20 mrad. 
For this measurement it 
 is however not sufficient to have only the main detector,
 which 
can accept electrons with an angle larger than 80 mrad only, since it leads
to too small event rates.

Although the measurement is still feasible for $100 \,\mbox{GeV}$, 
the energies of the tagged leptons should better go down 
to about $20 \,\mbox{GeV}$. 
In this region the \NLC\
provides an excellent possibility for testing the BFKL Pomeron.

\end{document}

%% file: gamma.pstex_t
\begin{picture}(0,0)%
\epsfig{file=gamma.pstex}%
\end{picture}%
\setlength{\unitlength}{0.00065600in}%
\begingroup\makeatletter\ifx\SetFigFont\undefined
% extract first six characters in \fmtname
\def\x#1#2#3#4#5#6#7\relax{\def\x{#1#2#3#4#5#6}}%
\expandafter\x\fmtname xxxxxx\relax \def\y{splain}%
\ifx\x\y   % LaTeX or SliTeX?
\gdef\SetFigFont#1#2#3{%
  \ifnum #1<17\tiny\else \ifnum #1<20\small\else
  \ifnum #1<24\normalsize\else \ifnum #1<29\large\else
  \ifnum #1<34\Large\else \ifnum #1<41\LARGE\else
     \huge\fi\fi\fi\fi\fi\fi
  \csname #3\endcsname}%
\else
\gdef\SetFigFont#1#2#3{\begingroup
  \count@#1\relax \ifnum 25<\count@\count@25\fi
  \def\x{\endgroup\@setsize\SetFigFont{#2pt}}%
  \expandafter\x
    \csname \romannumeral\the\count@ pt\expandafter\endcsname
    \csname @\romannumeral\the\count@ pt\endcsname
  \csname #3\endcsname}%
\fi
\fi\endgroup
\begin{picture}(3375,5520)(2971,-6709)
\put(6346,-2581){\makebox(0,0)[lb]{\smash{\SetFigFont{9}{10.8}{rm}$q$}}}
\put(6346,-3031){\makebox(0,0)[lb]{\smash{\SetFigFont{9}{10.8}{rm}$\bar{q}$}}}
\put(6346,-4831){\makebox(0,0)[lb]{\smash{\SetFigFont{9}{10.8}{rm}$q$}}}
\put(6346,-5281){\makebox(0,0)[lb]{\smash{\SetFigFont{9}{10.8}{rm}$\bar{q}$}}}
\put(4771,-5866){\makebox(0,0)[lb]{\smash{\SetFigFont{9}{10.8}{rm}$\gamma^*$}}}
\put(4771,-2041){\makebox(0,0)[lb]{\smash{\SetFigFont{9}{10.8}{rm}$\gamma^*$}}}
\put(4501,-6676){\makebox(0,0)[lb]{\smash{\SetFigFont{9}{10.8}{rm}${e^+}'$}}}
\put(4546,-1681){\makebox(0,0)[lb]{\smash{\SetFigFont{9}{10.8}{rm}$k_1'$}}}
\put(4501,-6181){\makebox(0,0)[lb]{\smash{\SetFigFont{9}{10.8}{rm}$k_2'$}}}
\put(4186,-2401){\makebox(0,0)[lb]{\smash{\SetFigFont{9}{10.8}{rm}$q_1$}}}
\put(4186,-5461){\makebox(0,0)[lb]{\smash{\SetFigFont{9}{10.8}{rm}$q_2$}}}
\put(2971,-5956){\makebox(0,0)[lb]{\smash{\SetFigFont{9}{10.8}{rm}$k_2$}}}
\put(4501,-1321){\makebox(0,0)[lb]{\smash{\SetFigFont{9}{10.8}{rm}${e^-}'$}}}
\put(3466,-6226){\makebox(0,0)[lb]{\smash{\SetFigFont{9}{10.8}{rm}$e^+$}}}
\put(3466,-1636){\makebox(0,0)[lb]{\smash{\SetFigFont{9}{10.8}{rm}$e^-$}}}
\put(3016,-1861){\makebox(0,0)[lb]{\smash{\SetFigFont{9}{10.8}{rm}$k_1$}}}
\end{picture}

%% file: fig2.pstex_t
\begin{picture}(0,0)%
\epsfig{file=fig2.pstex}%
\end{picture}%
\setlength{\unitlength}{0.00075000in}%
\begingroup\makeatletter\ifx\SetFigFont\undefined
% extract first six characters in \fmtname
\def\x#1#2#3#4#5#6#7\relax{\def\x{#1#2#3#4#5#6}}%
\expandafter\x\fmtname xxxxxx\relax \def\y{splain}%
\ifx\x\y   % LaTeX or SliTeX?
\gdef\SetFigFont#1#2#3{%
  \ifnum #1<17\tiny\else \ifnum #1<20\small\else
  \ifnum #1<24\normalsize\else \ifnum #1<29\large\else
  \ifnum #1<34\Large\else \ifnum #1<41\LARGE\else
     \huge\fi\fi\fi\fi\fi\fi
  \csname #3\endcsname}%
\else
\gdef\SetFigFont#1#2#3{\begingroup
  \count@#1\relax \ifnum 25<\count@\count@25\fi
  \def\x{\endgroup\@setsize\SetFigFont{#2pt}}%
  \expandafter\x
    \csname \romannumeral\the\count@ pt\expandafter\endcsname
    \csname @\romannumeral\the\count@ pt\endcsname
  \csname #3\endcsname}%
\fi
\fi\endgroup
\begin{picture}(9236,3811)(399,-3851)
\put(2950,-136){\makebox(0,0)[b]{\smash{\SetFigFont{8}{9.6}{rm}$y_1 y_2 \frac{d \sigma^{e^-e^+}}{dy_1 dy_2 dQ_1^2 dQ_2^2} \; [\mbox{pbarn}/\mbox{GeV}^4]$}}}
\put(2718,-3818){\makebox(0,0)[b]{\smash{\SetFigFont{9}{10.8}{rm}$Y = \log \frac{s}{s_0}$}}}
\put(1412,-811){\makebox(0,0)[lb]{\smash{\SetFigFont{10}{12.0}{rm}$Q^2 = 10 \,\mbox{GeV}^2$}}}
\put(7128,-137){\makebox(0,0)[b]{\smash{\SetFigFont{8}{9.6}{rm}$y_1 y_2 \frac{d \sigma^{e^-e^+}}{dy_1 dy_2 dQ_1^2 dQ_2^2} \; [\mbox{pbarn}/\mbox{GeV}^4]$}}}
\put(5217,-3495){\makebox(0,0)[rb]{\smash{\SetFigFont{9}{10.8}{rm}$10^{-6}$}}}
\put(4270,-3619){\makebox(0,0)[b]{\smash{\SetFigFont{9}{10.8}{rm}12}}}
\put(1166,-3619){\makebox(0,0)[b]{\smash{\SetFigFont{9}{10.8}{rm}2}}}
\put(1787,-3619){\makebox(0,0)[b]{\smash{\SetFigFont{9}{10.8}{rm}4}}}
\put(2408,-3619){\makebox(0,0)[b]{\smash{\SetFigFont{9}{10.8}{rm}6}}}
\put(3028,-3619){\makebox(0,0)[b]{\smash{\SetFigFont{9}{10.8}{rm}8}}}
\put(3649,-3619){\makebox(0,0)[b]{\smash{\SetFigFont{9}{10.8}{rm}10}}}
\put(5217,-2460){\makebox(0,0)[rb]{\smash{\SetFigFont{9}{10.8}{rm}$10^{-5}$}}}
\put(7774,-3619){\makebox(0,0)[b]{\smash{\SetFigFont{9}{10.8}{rm}8}}}
\put(8395,-3619){\makebox(0,0)[b]{\smash{\SetFigFont{9}{10.8}{rm}10}}}
\put(6843,-3818){\makebox(0,0)[b]{\smash{\SetFigFont{9}{10.8}{rm}$Y = \log \frac{s}{s_0}$}}}
\put(5589,-811){\makebox(0,0)[lb]{\smash{\SetFigFont{10}{12.0}{rm}$Q^2 = 25 \,\mbox{GeV}^2$}}}
\put(1092,-3495){\makebox(0,0)[rb]{\smash{\SetFigFont{9}{10.8}{rm}$10^{-5}$}}}
\put(7153,-3619){\makebox(0,0)[b]{\smash{\SetFigFont{9}{10.8}{rm}6}}}
\put(5217,-1426){\makebox(0,0)[rb]{\smash{\SetFigFont{9}{10.8}{rm}0.0001}}}
\put(5217,-391){\makebox(0,0)[rb]{\smash{\SetFigFont{9}{10.8}{rm}0.001}}}
\put(5291,-3619){\makebox(0,0)[b]{\smash{\SetFigFont{9}{10.8}{rm}0}}}
\put(5912,-3619){\makebox(0,0)[b]{\smash{\SetFigFont{9}{10.8}{rm}2}}}
\put(6533,-3619){\makebox(0,0)[b]{\smash{\SetFigFont{9}{10.8}{rm}4}}}
\put(1092,-391){\makebox(0,0)[rb]{\smash{\SetFigFont{9}{10.8}{rm}0.1}}}
\put(1092,-2719){\makebox(0,0)[rb]{\smash{\SetFigFont{9}{10.8}{rm}0.0001}}}
\put(1092,-1943){\makebox(0,0)[rb]{\smash{\SetFigFont{9}{10.8}{rm}0.001}}}
\put(1092,-1167){\makebox(0,0)[rb]{\smash{\SetFigFont{9}{10.8}{rm}0.01}}}
\end{picture}

%% file: fig3.pstex_t
\begin{picture}(0,0)%
\epsfig{file=fig3.pstex}%
\end{picture}%
\setlength{\unitlength}{0.00075000in}%
\begingroup\makeatletter\ifx\SetFigFont\undefined
% extract first six characters in \fmtname
\def\x#1#2#3#4#5#6#7\relax{\def\x{#1#2#3#4#5#6}}%
\expandafter\x\fmtname xxxxxx\relax \def\y{splain}%
\ifx\x\y   % LaTeX or SliTeX?
\gdef\SetFigFont#1#2#3{%
  \ifnum #1<17\tiny\else \ifnum #1<20\small\else
  \ifnum #1<24\normalsize\else \ifnum #1<29\large\else
  \ifnum #1<34\Large\else \ifnum #1<41\LARGE\else
     \huge\fi\fi\fi\fi\fi\fi
  \csname #3\endcsname}%
\else
\gdef\SetFigFont#1#2#3{\begingroup
  \count@#1\relax \ifnum 25<\count@\count@25\fi
  \def\x{\endgroup\@setsize\SetFigFont{#2pt}}%
  \expandafter\x
    \csname \romannumeral\the\count@ pt\expandafter\endcsname
    \csname @\romannumeral\the\count@ pt\endcsname
  \csname #3\endcsname}%
\fi
\fi\endgroup
\begin{picture}(3678,3855)(1155,-3632)
\put(3415,-3400){\makebox(0,0)[b]{\smash{\SetFigFont{9}{10.8}{rm}0.6}}}
\put(3760,-3400){\makebox(0,0)[b]{\smash{\SetFigFont{9}{10.8}{rm}0.7}}}
\put(4105,-3400){\makebox(0,0)[b]{\smash{\SetFigFont{9}{10.8}{rm}0.8}}}
\put(4450,-3400){\makebox(0,0)[b]{\smash{\SetFigFont{9}{10.8}{rm}0.9}}}
\put(4795,-3400){\makebox(0,0)[b]{\smash{\SetFigFont{9}{10.8}{rm}1}}}
\put(1788,115){\makebox(0,0)[b]{\smash{\SetFigFont{9}{10.8}{rm}events/year}}}
\put(3243,-3599){\makebox(0,0)[b]{\smash{\SetFigFont{9}{10.8}{rm}$y_1$}}}
\put(3071,-3400){\makebox(0,0)[b]{\smash{\SetFigFont{9}{10.8}{rm}0.5}}}
\put(1617,-3276){\makebox(0,0)[rb]{\smash{\SetFigFont{9}{10.8}{rm}0}}}
\put(1617,-3017){\makebox(0,0)[rb]{\smash{\SetFigFont{9}{10.8}{rm}20000}}}
\put(1617,-1983){\makebox(0,0)[rb]{\smash{\SetFigFont{9}{10.8}{rm}100000}}}
\put(1617,-689){\makebox(0,0)[rb]{\smash{\SetFigFont{9}{10.8}{rm}200000}}}
\put(2726,-3400){\makebox(0,0)[b]{\smash{\SetFigFont{9}{10.8}{rm}0.4}}}
\put(2381,-3400){\makebox(0,0)[b]{\smash{\SetFigFont{9}{10.8}{rm}0.3}}}
\put(2036,-3400){\makebox(0,0)[b]{\smash{\SetFigFont{9}{10.8}{rm}0.2}}}
\put(1691,-3400){\makebox(0,0)[b]{\smash{\SetFigFont{9}{10.8}{rm}0.1}}}
\end{picture}

%% file: fig4.pstex_t
\begin{picture}(0,0)%
\epsfig{file=fig4.pstex}%
\end{picture}%
\setlength{\unitlength}{0.00075000in}%
\begingroup\makeatletter\ifx\SetFigFont\undefined
% extract first six characters in \fmtname
\def\x#1#2#3#4#5#6#7\relax{\def\x{#1#2#3#4#5#6}}%
\expandafter\x\fmtname xxxxxx\relax \def\y{splain}%
\ifx\x\y   % LaTeX or SliTeX?
\gdef\SetFigFont#1#2#3{%
  \ifnum #1<17\tiny\else \ifnum #1<20\small\else
  \ifnum #1<24\normalsize\else \ifnum #1<29\large\else
  \ifnum #1<34\Large\else \ifnum #1<41\LARGE\else
     \huge\fi\fi\fi\fi\fi\fi
  \csname #3\endcsname}%
\else
\gdef\SetFigFont#1#2#3{\begingroup
  \count@#1\relax \ifnum 25<\count@\count@25\fi
  \def\x{\endgroup\@setsize\SetFigFont{#2pt}}%
  \expandafter\x
    \csname \romannumeral\the\count@ pt\expandafter\endcsname
    \csname @\romannumeral\the\count@ pt\endcsname
  \csname #3\endcsname}%
\fi
\fi\endgroup
\begin{picture}(7880,3867)(480,-3857)
\put(870,-2259){\makebox(0,0)[rb]{\smash{\SetFigFont{9}{10.8}{rm}40}}}
\put(870,-2570){\makebox(0,0)[rb]{\smash{\SetFigFont{9}{10.8}{rm}30}}}
\put(870,-2880){\makebox(0,0)[rb]{\smash{\SetFigFont{9}{10.8}{rm}20}}}
\put(870,-3191){\makebox(0,0)[rb]{\smash{\SetFigFont{9}{10.8}{rm}10}}}
\put(870,-3501){\makebox(0,0)[rb]{\smash{\SetFigFont{9}{10.8}{rm}0}}}
\put(870,-1949){\makebox(0,0)[rb]{\smash{\SetFigFont{9}{10.8}{rm}50}}}
\put(1634,-3625){\makebox(0,0)[b]{\smash{\SetFigFont{9}{10.8}{rm}0.3}}}
\put(1289,-3625){\makebox(0,0)[b]{\smash{\SetFigFont{9}{10.8}{rm}0.2}}}
\put(944,-3625){\makebox(0,0)[b]{\smash{\SetFigFont{9}{10.8}{rm}0.1}}}
\put(870,-397){\makebox(0,0)[rb]{\smash{\SetFigFont{9}{10.8}{rm}100}}}
\put(870,-1639){\makebox(0,0)[rb]{\smash{\SetFigFont{9}{10.8}{rm}60}}}
\put(870,-1328){\makebox(0,0)[rb]{\smash{\SetFigFont{9}{10.8}{rm}70}}}
\put(870,-1018){\makebox(0,0)[rb]{\smash{\SetFigFont{9}{10.8}{rm}80}}}
\put(870,-707){\makebox(0,0)[rb]{\smash{\SetFigFont{9}{10.8}{rm}90}}}
\put(1979,-3625){\makebox(0,0)[b]{\smash{\SetFigFont{9}{10.8}{rm}0.4}}}
\put(5563,-3625){\makebox(0,0)[b]{\smash{\SetFigFont{9}{10.8}{rm}0.2}}}
\put(5218,-3625){\makebox(0,0)[b]{\smash{\SetFigFont{9}{10.8}{rm}0.1}}}
\put(5144,-397){\makebox(0,0)[rb]{\smash{\SetFigFont{9}{10.8}{rm}0.24}}}
\put(5144,-656){\makebox(0,0)[rb]{\smash{\SetFigFont{9}{10.8}{rm}0.22}}}
\put(5144,-1690){\makebox(0,0)[rb]{\smash{\SetFigFont{9}{10.8}{rm}0.14}}}
\put(5144,-1432){\makebox(0,0)[rb]{\smash{\SetFigFont{9}{10.8}{rm}0.16}}}
\put(5144,-1173){\makebox(0,0)[rb]{\smash{\SetFigFont{9}{10.8}{rm}0.18}}}
\put(5144,-914){\makebox(0,0)[rb]{\smash{\SetFigFont{9}{10.8}{rm}0.2}}}
\put(5908,-3625){\makebox(0,0)[b]{\smash{\SetFigFont{9}{10.8}{rm}0.3}}}
\put(6770,-3824){\makebox(0,0)[b]{\smash{\SetFigFont{9}{10.8}{rm}$y_1$}}}
\put(8322,-3625){\makebox(0,0)[b]{\smash{\SetFigFont{9}{10.8}{rm}1}}}
\put(7977,-3625){\makebox(0,0)[b]{\smash{\SetFigFont{9}{10.8}{rm}0.9}}}
\put(7632,-3625){\makebox(0,0)[b]{\smash{\SetFigFont{9}{10.8}{rm}0.8}}}
\put(6253,-3625){\makebox(0,0)[b]{\smash{\SetFigFont{9}{10.8}{rm}0.4}}}
\put(6598,-3625){\makebox(0,0)[b]{\smash{\SetFigFont{9}{10.8}{rm}0.5}}}
\put(6942,-3625){\makebox(0,0)[b]{\smash{\SetFigFont{9}{10.8}{rm}0.6}}}
\put(7287,-3625){\makebox(0,0)[b]{\smash{\SetFigFont{9}{10.8}{rm}0.7}}}
\put(5144,-1949){\makebox(0,0)[rb]{\smash{\SetFigFont{9}{10.8}{rm}0.12}}}
\put(1488,-110){\makebox(0,0)[b]{\smash{\SetFigFont{9}{10.8}{rm}events/$1000 \mbox{pb}^{-1}$}}}
\put(2496,-3824){\makebox(0,0)[b]{\smash{\SetFigFont{9}{10.8}{rm}$y_1$}}}
\put(4048,-3625){\makebox(0,0)[b]{\smash{\SetFigFont{9}{10.8}{rm}1}}}
\put(3703,-3625){\makebox(0,0)[b]{\smash{\SetFigFont{9}{10.8}{rm}0.9}}}
\put(2324,-3625){\makebox(0,0)[b]{\smash{\SetFigFont{9}{10.8}{rm}0.5}}}
\put(2668,-3625){\makebox(0,0)[b]{\smash{\SetFigFont{9}{10.8}{rm}0.6}}}
\put(3013,-3625){\makebox(0,0)[b]{\smash{\SetFigFont{9}{10.8}{rm}0.7}}}
\put(3358,-3625){\makebox(0,0)[b]{\smash{\SetFigFont{9}{10.8}{rm}0.8}}}
\put(1426,-1261){\makebox(0,0)[lb]{\smash{\SetFigFont{10}{12.0}{rm}$\theta_{tag}> 30 \,\mbox{mrad}$}}}
\put(5144,-2208){\makebox(0,0)[rb]{\smash{\SetFigFont{9}{10.8}{rm}0.1}}}
\put(5144,-2466){\makebox(0,0)[rb]{\smash{\SetFigFont{9}{10.8}{rm}0.08}}}
\put(5144,-2725){\makebox(0,0)[rb]{\smash{\SetFigFont{9}{10.8}{rm}0.06}}}
\put(5144,-2984){\makebox(0,0)[rb]{\smash{\SetFigFont{9}{10.8}{rm}0.04}}}
\put(5701,-1261){\makebox(0,0)[lb]{\smash{\SetFigFont{10}{12.0}{rm}$\theta_{tag}> 80 \,\mbox{mrad}$}}}
\put(5763,-110){\makebox(0,0)[b]{\smash{\SetFigFont{9}{10.8}{rm}events/$1000 \mbox{pb}^{-1}$}}}
\put(5144,-3501){\makebox(0,0)[rb]{\smash{\SetFigFont{9}{10.8}{rm}0}}}
\put(5144,-3242){\makebox(0,0)[rb]{\smash{\SetFigFont{9}{10.8}{rm}0.02}}}
\end{picture}